%% file: main.tex
\documentclass[10pt,twocolumn,letterpaper]{article}

\usepackage[accsupp]{axessibility}

\usepackage{iccv}
\usepackage{times}
\usepackage{epsfig}
\usepackage{graphicx}
\usepackage{amsmath}
\usepackage{amssymb}

\usepackage{caption}
\usepackage{subcaption}
\usepackage{multirow}
\usepackage{comment}

\usepackage[breaklinks=true,bookmarks=false]{hyperref}

\iccvfinalcopy %

\ificcvfinal\fi

\input{commands}

\begin{document}

\title{Surface Extraction from Neural Unsigned Distance Fields}

\author{Congyi Zhang$^{1,2*}$ \and
Guying Lin$^{1*}$ \and
Lei Yang$^{1,2}$ \and
Xin Li$^3$ \and
Taku Komura$^1$ \and
Scott Schaefer$^3$ \and
John Keyser$^3$ \and
Wenping Wang$^{3\dagger}$ \and
\\
$^1$The University of Hong Kong \qquad
$^2$TransGP \qquad \\
$^3$Texas A\&M University
\footnotetext{Equal contribution}
}

\maketitle
\ificcvfinal\thispagestyle{empty}\fi
\def\thefootnote{*}\footnotetext{Equal contribution}
\def\thefootnote{$\dagger$}\footnotetext{Corresponding author}
\def\thefootnote{1}
\begin{abstract} 

We propose a method, named DualMesh-UDF, to extract a surface from unsigned distance functions (UDFs), encoded by neural networks, or neural UDFs. Neural UDFs are becoming increasingly popular for surface representation because of their versatility in presenting surfaces with arbitrary topologies, as opposed to the signed distance function that is limited to representing a closed surface.
However, the applications of neural UDFs are hindered by the notorious difficulty in extracting the target surfaces they represent. Recent methods for surface extraction from a neural UDF suffer from significant geometric errors or topological artifacts due to two main difficulties: (1) A UDF does not exhibit sign changes; and (2) A neural UDF typically has substantial approximation errors.

DualMesh-UDF addresses these two difficulties. Specifically, given a neural UDF encoding a target surface $\bar{S}$ to be recovered, we first estimate the tangent planes of $\bar{S}$ at a set of sample points close to $\bar{S}$. Next, we organize these sample points into local clusters, and for each local cluster, solve a linear least squares problem to determine a final surface point. These surface points are then connected to create the output mesh surface, which approximates the target surface. The robust estimation of the tangent planes of the target surface and the subsequent minimization problem constitute our core strategy, which  contributes to the favorable performance of DualMesh-UDF over other competing methods. To efficiently implement this strategy, we employ an adaptive Octree. Within this framework, we estimate the location of a surface point in each of the octree cells identified as containing part of the target surface. Extensive experiments show that our method outperforms existing methods in terms of surface reconstruction quality while maintaining comparable computational efficiency.

\end{abstract}

\input{sections/introduction}
\input{sections/related}

\input{sections/preliminary}
\input{sections/method}
\input{sections/results}
\input{sections/conclusion}

{\small
\bibliographystyle{ieee_fullname}
\bibliography{egbib}
}

\clearpage
\appendix
\twocolumn[
    \centering
    \Large
    \textbf{Surface Extraction from Neural Unsigned Distance Fields} \\
    \vspace{1.0em}
    Supplementary Material\\
    \vspace{2.0em}
]

\input{supp_sections/MLP_characteristics}
\input{supp_sections/training_settings}
\input{supp_sections/experiments}

\end{document}

%% file: commands.tex
\newcommand{\normal}{\mathbf{n}}

\newcommand{\point}{\mathbf{p}}

\newcommand{\xpoint}{\mathbf{x}}

\newcommand{\vpoint}{\mathbf{v}}

\newcommand{\surface}{\bar {\mathcal S}}
\newcommand{\udf}{\bar F}
\newcommand{\diag}{\mathrm{Diag}}

\newcommand{\centroid}{\mathbf{c}}

\newcommand{\proj}{\mathbf{q}}

\DeclareMathOperator*{\argmin}{arg\,min}

%% file: sections/introduction.tex
\section{Introduction}

Implicit surfaces are widely used for surface representation in computer vision and computer graphics. 
An implicit surface is usually defined as a level set of a function, such as the zero-level set of a signed distance function (SDF). Extracting a mesh representation of an implicit surface from its defining equation is therefore a critical task for surface visualization and processing. Recent advances in machine learning have given rise to a new kind of implicit surface, called a {\em neural implicit surface}. A neural implicit surface is a level-set of a function encoded by an MLP (multilayer perceptron) and has the advantage of compactness and inherent smoothness thanks to its MLP representation. SDFs or occupancy fields are widely used in these implicit representations~\cite{Park2019DeepSDF, Mescheder2019OccNet, Michalkiewicz2019ISR, Chibane2020Implicit, Sitzmann2020SIREN, Tancik2020PE, Gropp2020Implicit, Atzmon2021SALD, Jiang2020Local, Martel2021Acorn, Takikawa2021NGLOD}.

However, neural implicit surfaces based on the SDF or occupancy fields require inside/outside labeling and thus can only represent orientable and closed surfaces. Hence, as an extension, unsigned distance functions (UDFs) have been used to represent surfaces of arbitrary topologies, including open surfaces with boundaries or non-orientable surfaces (\eg the M\"{o}bius strip).
Despite its versatility, applications of a UDF-based surface representation are severely hindered by the difficulty in extracting the target surface it represents, as shown in \cite{Chibane2020UDF} and~\cite{Guillard2022MeshUDF}. 

{\bf Problem formulation:} 
Suppose that a surface $\surface$, called the {\em target surface}, is defined as the zero-level set of its unsigned distance function (UDF) $\udf(\point)$.
Then suppose that this UDF $\udf(\point)$ is approximated by a neural network with the resulting network-encoded UDF being referred to as the {\em neural UDF}, denoted by $F(\point)$.  Given a neural UDF $F(\point)$, the surface extraction problem is to robustly extract a surface $\mathcal{S}$ from $F(\point)$ such that $\mathcal{S}$ well approximates the target surface $\surface$.

\textbf{Challenges:} 
The difficulty in surface extraction from a neural UDF arises from two aspects:  
(1) A UDF does not have zero-crossings (or sign changes) across the surface it represents. As a result, traditional mesh extraction methods that rely on zero-crossings (\eg Marching Cubes~\cite{Lorensen1987MC,Kobbelt2001EMC}, Dual Contouring~\cite{Ju2002DC}, and their variants) are not applicable to UDFs.
(2) The MLP representation of a neural UDF tends to have significant approximation errors around the target surface (see detailed error characteristics of neural UDFs in Sec.~\ref{sec:errors}). This makes it even more challenging to extract a high-quality approximation of the target surface. 

Several methods,  MeshUDF~\cite{Guillard2022MeshUDF}, CAP-UDF~\cite{Zhou2022UDF} and Neural Dual Contouring (NDC)~\cite{Chen2022NDC}, have recently been proposed for extracting a mesh surface from a UDF. MeshUDF and CAP-UDF attempt to infer the gradients of a UDF on the grids and determine the sign changes of the estimated gradients, invoking the Marching Cubes method for mesh extraction. When applied to a neural UDF, the sign-change inference step of this method suffers from instability due to the non-negligible error introduced by the approximate MLP representation near the surface where the gradients of the ideal UDF are undefined. 
As a result, the extracted meshes are less accurate and often have topological errors (\eg holes).
The NDC method proposes a data-driven Dual Contouring approach to predict the position of mesh vertices and dual faces directly from the UDF data. When applied to a neural UDF, this method often produces meshes with considerable artifacts such as holes, zig-zags, etc.

We develop a new strategy, consisting of novel sampling and efficient optimization techniques to address the difficulties in surface extraction from the neural UDF. Suppose the input is a neural UDF $F(\point)$ encoding the target surface $\surface$ to be recovered. Our strategy has two key steps: (1) {\em computing approximate tangent planes of the target surface}; and (2) {\em local minimization for generating final surface points}. 

In Step (1), we first generate sample points $\mathbf{p}_i$ around, but not too close to, the target surface $\surface$, because the UDF values and gradients at locations too close to $\surface$ are relatively unreliable. Thus, $\mathbf{p}_i$ are called {\em off-surface sample points.} For each $\mathbf{p}_i$, we use the UDF value $F(\mathbf{p}_i)$ and its gradient $\nabla F(\mathbf{p}_i)$ to project $\mathbf{p}_i$ towards the target surface to obtain point $\mathbf{q}_i = \mathbf{p}_i - F(\mathbf{p}_i) \mathbf{n}_i(\mathbf{p}_i)$ where $\mathbf{n}(\mathbf{p}_i) = \nabla F(\mathbf{p}_i) / \|\nabla F(\mathbf{p}_i)\|$~\cite{Chibane2020UDF}. These points $\mathbf{q}_i$ are called {\em projection points}. Although the points $\mathbf{q}_i$ are very close to the target surface $\surface$, as we will show later, the noisy error in the neural UDF makes these points a poor approximation to the target surface.

To further improve surface accuracy, for each projection point $\mathbf{q}_i$ we generate an estimated tangent plane $T_i$ of $\surface$ such that $T_i$ passes through $\mathbf{q}_i$ and has the unit normal vector $\mathbf{n}_i$.  Note that the normal vector $\mathbf{n}_i$ of $T_i$ is set to be $\mathbf{n}(\mathbf{p}_i)$ rather than $\mathbf{n}(\mathbf{q}_i)$ since the former is a more reliable estimation.  This is because the initial sample point $\mathbf{p}_i$ is not too close to $\surface$, so the gradient $\nabla F(\mathbf{p}_i)$ is less contaminated by the pronounced errors of the neural UDF close to $\surface$. 

In Step (2), the estimated tangent planes are organized into clusters, which may overlap. For each cluster of tangent planes $T_i$, we solve a linear least squares problem to produce a final surface point $\mathbf{s}_i$ that minimizes the sum of its squared distances to the tangent planes $T_i$. This minimization step based on tangent planes not only provides an accurate surface point but also allows us to faithfully reconstruct the sharp edges of the target surface. Finally, all the surface points $\mathbf{s}_i$ from all the clusters are connected to form the output mesh surface to approximate the target surface $\surface$.

To efficiently implement the above strategy, our {\em DualMesh-UDF} method adopts an adaptive Octree structure to partition the space containing the target surface to regular cells. We developed efficient procedures to determine those cells that contain part of the target surface and perform the sampling and minimization procedures in each occupied cell. 
To connect the surface points to create the output mesh, we follow the Dual Contouring approach, connecting surface points residing in adjacent grid cells to create polygons dual to octree edges. 

Extensive experiments demonstrate that our DualMesh-UDF significantly outperforms existing methods in terms of surface reconstruction accuracy and sharp feature preservation.

The main contribution of this work is a new algorithm to robustly and accurately extract a surface from a neural UDF. To overcome the inevitable approximation errors near the target surface and cut locus, we obtain robust estimation of surface tangent planes by leveraging off-surface sample points, use least square minimization to better predict the surface points, and achieve high-quality surface extraction results with sharp features better preserved compared to the state of the art. The code is available at \href{https://github.com/cong-yi/DualMesh-UDF}{https://github.com/cong-yi/DualMesh-UDF}.

%% file: sections/related.tex
\section{Related Work}

The Marching Cubes method \cite{Lorensen1987MC} and its variants have been established as the \emph{de facto} standard of converting distance fields to boundary mesh representations. 
Using the gradient information of the distance function, Extended Marching Cubes~\cite{Kobbelt2001EMC} and Dual Contouring \cite{Ju2002DC} are both capable of producing meshes with faithfully preserved sharp features (\eg, corners and edges).

However, all of these methods require inside/outside labeling on the sampling grid, which is either a regular grid of cubes or an adaptive octree grid of cells, to determine whether any zero-level set surface of the signed distance field crosses a particular grid cell. 
This requirement for sign changes limits the application of these methods to only SDFs or its variants that have sign changes across the underlying surface. 
Hence, they cannot be directly used to extract a mesh surface from a UDF, which is now often used to represent arbitrary surfaces such as open surfaces or non-orientable surfaces~\cite{Chibane2020UDF}. 

Our method is similar to \cite{Ju2002DC} only in the sense that we follow \cite{Ju2002DC} to solve a quadratic error function (QEF) to estimate a surface point per grid cell. But unlike \cite{Ju2002DC}, where sign changes are available and the gradients at the zero-crossings are reliable, unsigned distance fields do not have sign changes and are non-differentiable at the zero level set. Additionally, we will show that the information around the zero-level set of a neural UDF is unreliable. Due to these two reasons, we alternatively make use of the spatial sample points that are \emph{off} the target surface in the neural UDF to find reliably estimated tangent planes and formulate a QEF for estimating the final surface points.

Three relevant methods, MeshUDF~\cite{Guillard2022MeshUDF}, CAP-UDF~\cite{Zhou2022UDF} and Neural Dual Contouring (NDC)~\cite{Chen2022NDC}, have recently been proposed to extract meshes from UDFs.  
MeshUDF~\cite{Guillard2022MeshUDF} and CAP-UDF~\cite{Zhou2022UDF} use the gradient information of the UDF to assign different signs ($+$ or $-$) to grid points on different sides of the underlying surface, thus invoking Marching Cubes to extract the surface according to the sign labels. 
However, the quality of the estimated sign labels can be significantly affected by the accuracy of the neural UDF. As we will show later, the neural UDF becomes less accurate around the target surface it represents, contaminating the inferred sign labels and explaining the poor performance of these two methods.
Their use of Marching Cubes is incapable of preserving sharp corners or edges, while the use of a regular grid incurs high memory overhead when a high grid resolution is needed to resolve shape details.
In contrast, our DualMesh-UDF method preserves sharp features and uses an adaptive octree grid to reduce computational expense even with a high grid resolution. Furthermore, estimating the surface point location is done by solving a least square problem (QEF) which is more robust than the gradient-based sign-labeling strategy used in MeshUDF in terms of the topology of the resulting meshes.

The NDC method~\cite{Chen2022NDC} uses a data-driven approach to train a neural network that predicts vertex position per regular grid cell and the overall dual faces directly from a UDF. 
However, as a data-driven method, NDC's performance is largely influenced by how accurately the network is trained. 
We show in the extensive experiments that our explicit geometry-based method consistently outperforms NDC in terms of reconstruction accuracy, preserving sharp features, and preserving smooth boundaries in the original shapes.

%% file: sections/preliminary.tex
\section{Error Characteristics of Neural UDF} \label{sec:errors}

We first analyze the characteristics of the errors introduced by the MLP representation used to approximate a UDF, which explains why extracting a mesh surface from a neural UDF is difficult. 
This analysis also provides a foundation for justifying design choices in our method to overcome these challenging characteristics and achieve robust surface extraction.

We begin with an ideal UDF $\bar F(\xpoint)$ that represents a target surface $\surface$ defined by the zero-level set of $\bar F(\xpoint)$, that is, $\surface = \{ \xpoint | \bar F(\xpoint) = 0\} \subset \mathbb{R}^3$. Note that the UDF $\bar F(\xpoint)$ is non-negative, so it does not have sign changes across the surface $\surface$.
Furthermore, $\bar F(\xpoint)$ is not differentiable at the surface $\surface$ or the surface's cut locus~\footnote{The cut locus of a surface is a set of points such that each point of the set has two or more distinct closest points on the surface.}~\cite{Jones2006Survey}.

\begin{figure}
    \centering
    \includegraphics[width=\linewidth]{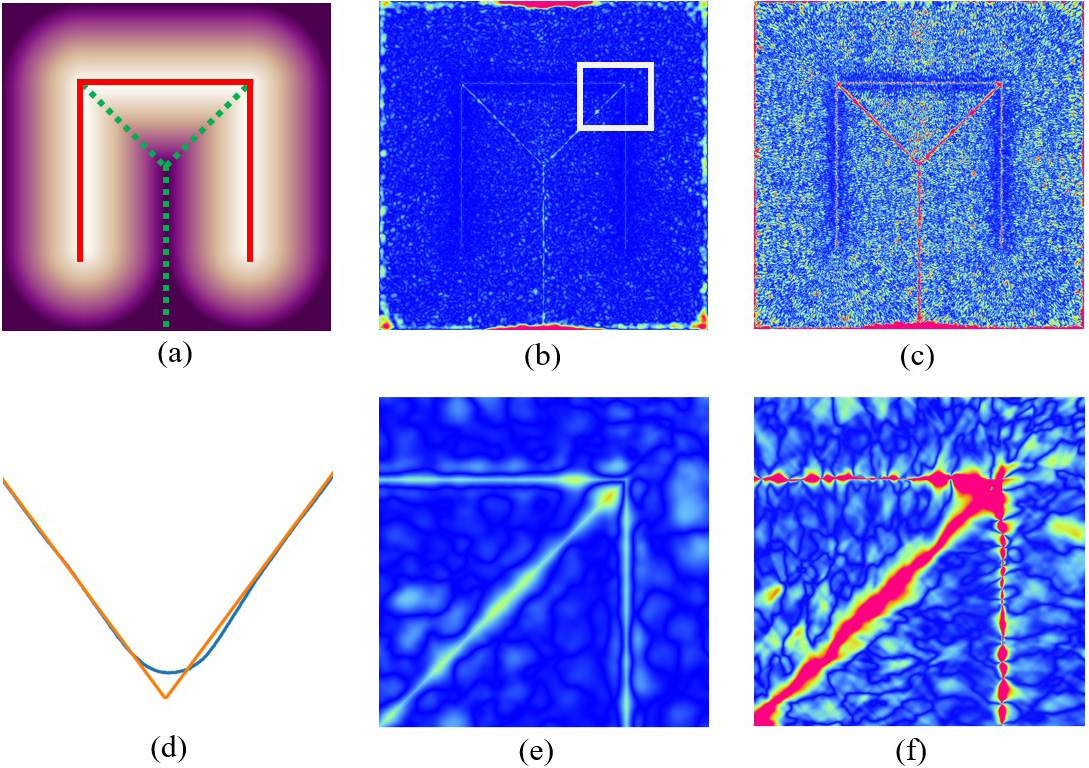}
    \caption{A 2D illustrative example. We observe that the neural UDF tends to have larger errors near the zero-level set $\surface$ and its cut locus. (a) The target shape (the red solid curve), its cut locus (green dashed), and the induced GT UDF;
    (b) Approximation errors of the neural UDF to the GT UDF; (c) Gradient direction errors between the neural UDF and the GT UDF;
    For the local region outlined with the white box around the corner, we show the close-up views of the distance error and the gradient direction error of the region in (e) and (f), respectively;
    (d) The GT UDF and its approximation by a neural UDF showing that the neural UDF is positive and smooth at the zero-level set of the ideal UDF, but the ideal UDF is non-differentiable.
    }
    \label{fig:error_analysis}
\end{figure}

Now suppose that the ideal UDF $\bar F(\xpoint)$ is approximated by an MLP, denoted by $F(\xpoint)$. The approximation errors of $F(\xpoint)$ to $\bar F(\xpoint)$ are significant in the narrow region around the surface and a narrow region around the cut locus of $\surface$, because the neural UDF $F(\xpoint)$ is inherently smooth and thus poorly approximates the ideal $\bar F(\xpoint)$, which is non-smooth at the surface $\surface$ and its cut locus. Due to these errors, in general, we have $F(\xpoint) > 0$ for any $\xpoint \in \surface$ when the MLP is differentiable.

For ease of visualization, we will use a 2D example (see Fig.~\ref{fig:error_analysis}) to illustrate the characteristic behaviors of the approximation errors of $F(\xpoint)$, whose zero-level set defines an open curve (red in Fig.~\ref{fig:error_analysis}(a)). 
The characteristics of the approximation errors of $F(\xpoint)$ in 3D space are similar. Detailed visualization and analysis of the error behaviors in 3D space are provided in the supplementary materials. 
Note that the cut locus (Fig.~\ref{fig:error_analysis}(a)) of $\surface$ touches the sharp features of $\surface$. Therefore the neural UDF $F(\xpoint)$ has even more significant errors in terms of both the distance value and the gradient direction around the sharp features (see the error maps of the distance values and the gradient directions in Fig.~\ref{fig:error_analysis}(e,f), respectively). Consequently, in 3D space, the resulting inaccurate distance values and unreliable gradient vectors of the neural UDF $F(\xpoint)$ make it hard to faithfully reconstruct the target surface, especially to preserve the sharp edges in the extracted surface. 

To recap, given a neural UDF $F(\xpoint)$, there are mainly two reasons for the difficulty in extracting a surface $\surface_M$ to approximate the target surface $\surface$ defined by the ideal UDF $\bar F(\xpoint)$: (1) the given neural UDF $F(\xpoint)$ is usually a poor approximation of the ideal UDF $\bar F(\xpoint)$ around the target surface $\bar {\mathcal S}$ and its cut locus, where $\bar F({\bf x})$ is non-differentiable; and (2) the neural UDF $F({\bf x})$ does not, in general, have a zero-crossing around the target surface $\surface$, thus no well-defined surface is associated with it. 
These issues, faced by current deep neural networks, make it hard to estimate the location of the surface in a numerically stable manner and thus motivate us to develop the two filtering criteria described in Sec.~\ref{sec:adaptation}.

%% file: sections/method.tex
\section{Method}
Our method consists of three major designs: 
(1) a quadratic error function (QEF) that solves for a surface point within a cell for mesh extraction (Sec.~\ref{sec:qef}); (2) an adaptive octree data structure that produces high-resolution grid cells while reducing the number of QEF solves required 
(Sec.~\ref{sec:octree}); and 
(3) a point filtering strategy (Sec.~\ref{sec:adaptation}) for a neural UDF, whose distance values and gradient directions are considerably more noisy compared to the GT as discussed in Sec.~\ref{sec:errors}.

The proposed DualMesh-UDF pipeline has three main steps. Firstly, it employs an adaptive tree to partition the domain of the UDF containing the target surface $\surface$. 
Then, it detects non-empty cells (defined in Sec.~\ref{sec:octree}) based on a two-step cell-shape intersection detection method, and solves one surface point per non-empty leaf cell in a least squares manner. 
For a neural network, two distance bounds are introduced to filter out sample points that may introduce noisy information and contaminate the localization of the target surface. 
Lastly, since there are no sign changes in the UDFs, we construct the mesh faces by checking each edge shared by four incident non-empty cells in the octree. 

\subsection{QEF to locate surface points}\label{sec:qef}

\begin{figure}
\centering
    \begin{subfigure}{0.49\linewidth}
    \captionsetup{justification=centering}
    \includegraphics[width=\textwidth]{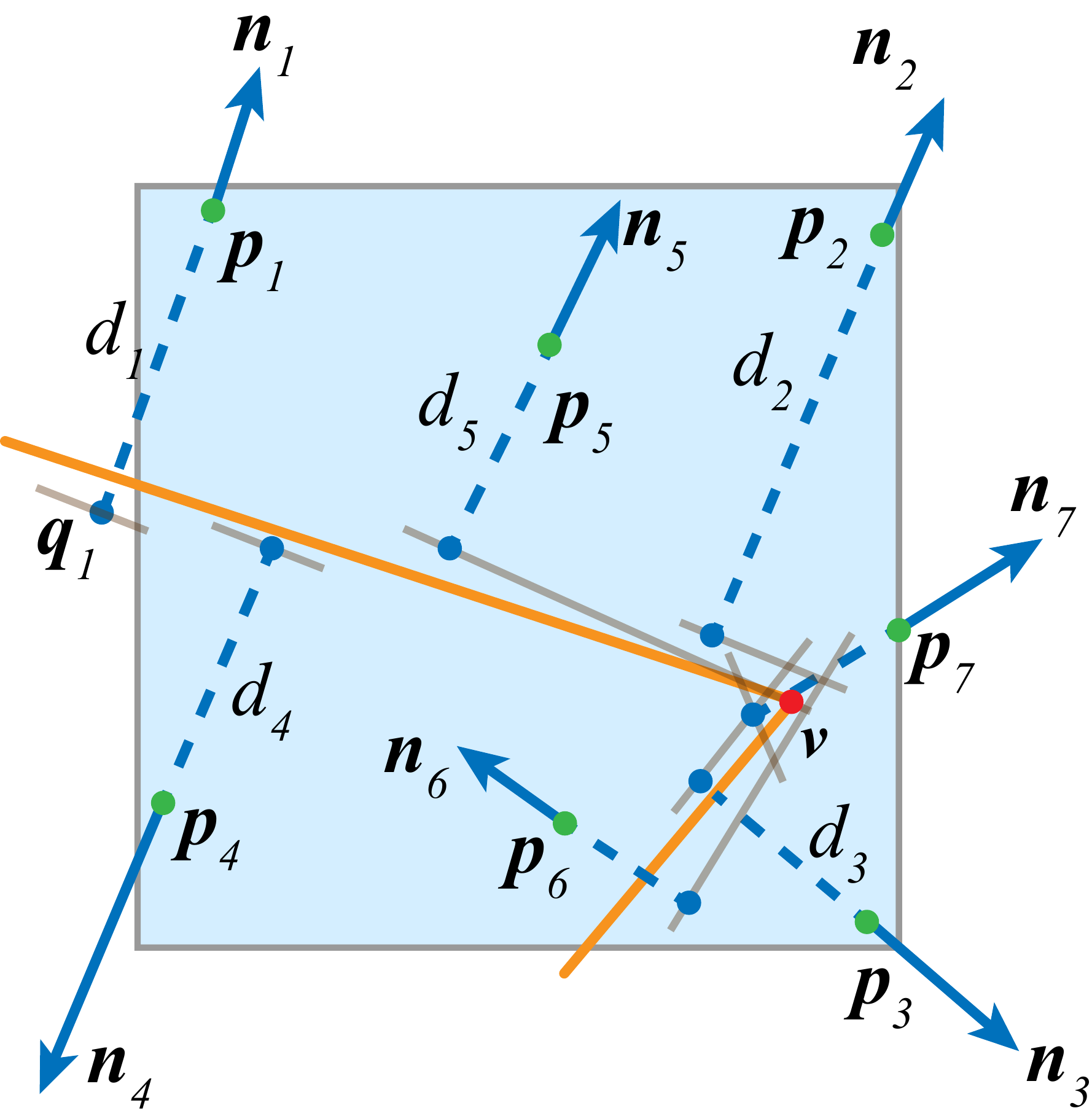}
    \caption{A neural UDF case}
    \label{fig:our_method_for_udf_noisy}
  \end{subfigure}
  \begin{subfigure}{0.49\linewidth}
  \captionsetup{justification=centering}
    \includegraphics[width=\textwidth]{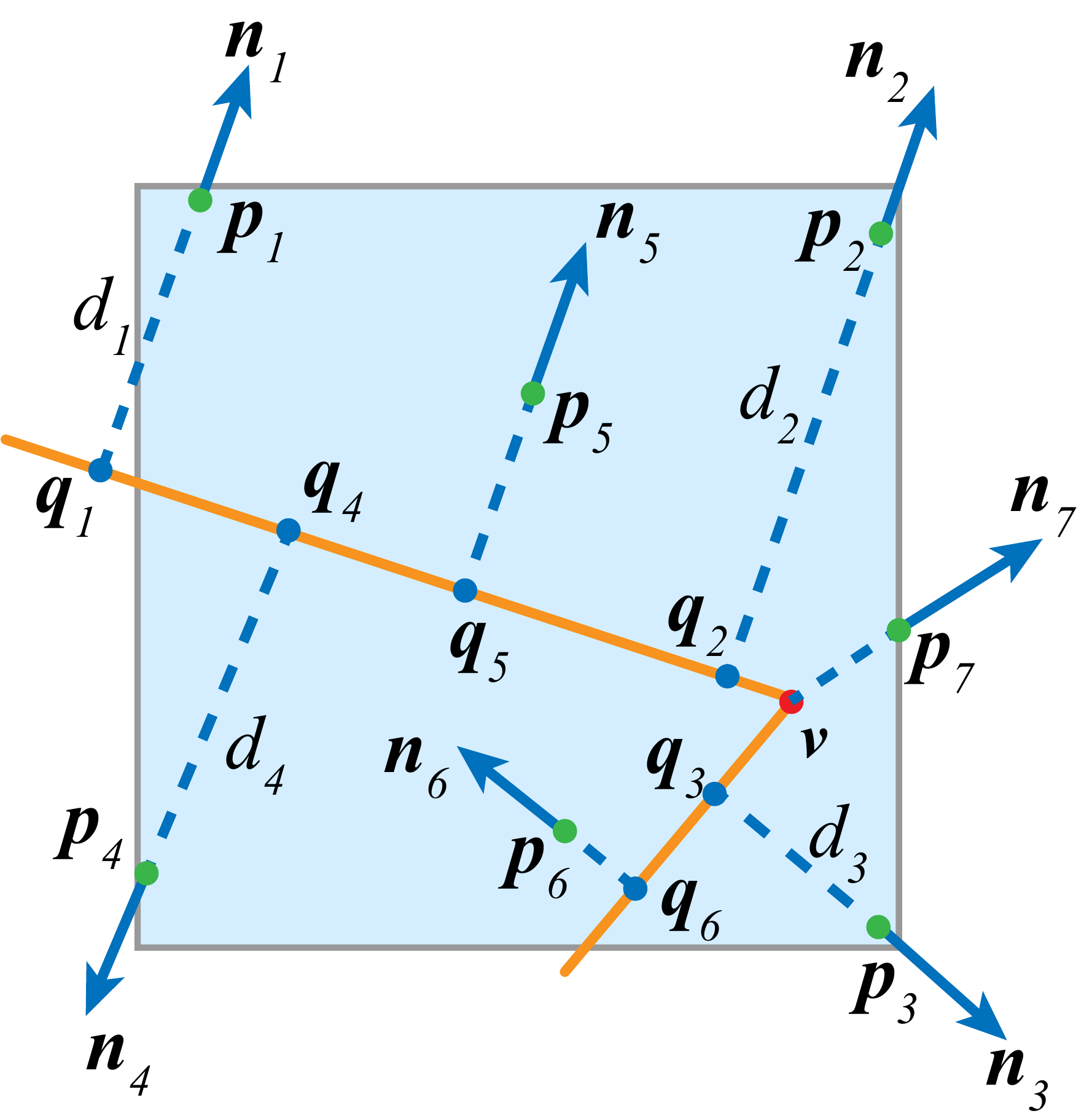}
    \caption{An ideal UDF case}
    \label{fig:our_method_for_gt_udf}
  \end{subfigure}\hfill

  \caption{The QEF formulation for the unsigned distance field. The orange lines represent the target surface to be reconstructed. The gray lines in (a) stand for the estimated tangent planes contributed by the corresponding sample points. The red point is the solution to the QEF problem.}
  \label{fig:method_illustration}
\end{figure}

Given a cell that contains part of the target surface, we present a procedure to estimate a surface point in this non-empty cell, as illustrated in Fig.~\ref{fig:method_illustration} for 2D demonstrations.
For a neural UDF, we aim to compute a reconstructed point per non-empty cell. For a set of sample points $\{\point_i\}_{i=1}^m$ in the given cell, we query the neural UDF $F(\xpoint)$ to get the approximate distance value and the normalized gradient vector\footnote{While $\|\nabla F\|$ should be 1 for a distance function, note that since $\nabla F$ is drawn from a neural representation, it may differ slightly, and thus need normalization.} at $\point_i$, \ie, $d_i = F(\point_i)$ and $\normal_i = \nabla F(\point_i)/\| \nabla F(\point_i) \|$, respectively (see Fig.~\ref{fig:our_method_for_udf_noisy}).

Each of the points contributes to an estimated tangent plane of the target surface $\bar {\mathcal S}$: $\normal_i\cdot(\point_i-\xpoint)-d_i = 0$. 
Considering the inaccurate nature of the neural UDF, we define the estimated surface point to be the point that is the closest to all these tangent planes and compute it by minimizing a \emph{quadratic error function} (a least squares problem): 
\begin{equation}\label{eq:qef}
    \vpoint=\argmin_{\xpoint}\sum_{i=1}^m\left(\normal_i\cdot(\point_i-\xpoint)-d_i\right)^2,
\end{equation}
where $m$ is the number of points $\{\point_i\}$ and $\xpoint$ is a surface point to be solved. 
In our implementation, we solve the linear least squares problem in Eqn.~\ref{eq:qef} of the form $\min\|Ax-b\|^2$.
For a non-empty cell, this procedure can yield a reconstructed point $\vpoint$, which the target surface $\surface$ approximately crosses.

However, when $A$ is nearly singular, the solution to Eqn.~\ref{eq:qef} can be located near the boundary of the cell or even outside of it, which would compromise the quality of extracted mesh surfaces.
Note that while \cite{Ju2002DC} proposed an approach with an additional regularization point to stabilize such cases, their approach relies on the intersection points at the zero-level set of an SDF with the cell edges. However, in our setting, the zero-level set of the UDF is elusive or unavailable, therefore we do not have such intersections. We propose an alternative approach based on singular value analysis of the matrix $A$.
We denote the three singular values of the matrix $A$ as $\sigma_0\geq\sigma_1\geq\sigma_2\geq0$ and consider three cases:

1) In the non-singular case, where all three singular values are much larger than $0$, $\vpoint$ corresponds to a \emph{sharp feature point} that can be solved directly from Eqn.~\ref{eq:qef}.

2) If only $\sigma_2 \approx 0$, the solution space of $\vpoint$ corresponds to a linear edge within the cell. This line has the same direction with the singular vector corresponding to $\sigma_2$ and passes the solution of Eqn.~\ref{eq:qef}. We then compute the intersections between this line and all 6 faces of the cell, yielding two intersecting points. We set $\vpoint$ as the midpoint of these two points.

3) In the case where both $\sigma_{1,2} \approx0$, the part of the target surface enclosed in this cell is approximately planar. Similar to the edge case, we can formulate a plane function from the corresponding singular vectors. By computing the intersecting points between all 12 edges of the cell and this plane, we obtain multiple intersecting points. We use the centroid of these points as $\vpoint$.

This way, in the degenerate cases (2 and 3), the point $\vpoint$ is still a good approximate solution to Eqn.~\ref{eq:qef} and well positioned inside the cell.

In a neural UDF (\eg, Fig.~\ref{fig:our_method_for_udf_noisy}), the UDF values $\{d_i\}$ and the gradient directions $\{\normal_i\}$ are approximated and often unreliable. 
Consequently, the tangent plane computed from each sample point $\point_i$ may not exactly align with the target surface.
Solving Eqn.~\ref{eq:qef} yields a least squares solution that is numerically robust to the approximation errors in neural UDF and can lead to an accurate estimation of the surface point in this non-empty cell. For an ideal UDF case, our method is also applicable to yield an accurate result as shown in Fig.~\ref{fig:our_method_for_gt_udf}.

Furthermore, to enhance the robustness of our method, we also design a point filtering strategy to remove unreliable sample points, especially those near the target surface and its cut locus, where larger approximation errors exist in the MLP-encoded UDF. The details of this filtering strategy are explained in Sec.~\ref{sec:adaptation}.

\textbf{Differences to DC~\cite{Ju2002DC}.}
Note that the dual contouring (DC) method \cite{Ju2002DC} designs a QEF using intersection points between the surface and cell edges, along with the gradient directions at these intersection points. This approach is not applicable to our objective of extracting surface meshes from a neural UDF. This is because 
(1) it is difficult to find a reliable intersection point, as an exact zero-level set that crosses an edge may not exist, and
(2) the gradient directions in the region with lower UDF values (\ie, closer to the target surface) are highly unstable and thus unreliable for estimating the tangent plane. 
Hence, our method leverages \textit{off-surface} points that are sufficiently far from the target surface, as indicated by their UDF values (see Sec.~\ref{sec:adaptation}). 
Our approach thus has less stringent requirements than the DC method (as presented in \cite{Ju2002DC}), yet produces satisfactory results even on noisy neural UDFs as demonstrated by our results.

\subsection{Octree Design and Subdivision}
\label{sec:octree}

\begin{figure}
    \centering
    \includegraphics[width=\linewidth]{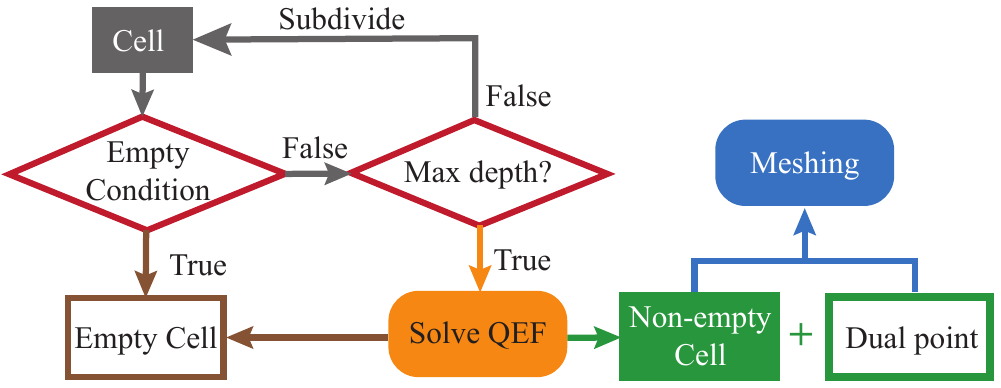}
    \caption{The pipeline of our mesh extraction method.}
    \label{fig:pipeline}
\end{figure}

To efficiently process high-resolution data and speed up the mesh extraction process, we employ an octree data structure.
As shown in Fig.~\ref{fig:pipeline}, we recursively subdivide the cells unless they are categorized as \textit{empty} according to the following criterion or they reach the maximal depth. 
After the subdivision, all leaf nodes that are non-empty will invoke the QEF procedure to solve for the dual points.

\textbf{Checking obviously empty cells.}
Firstly, we design a checking condition to help quickly prune obviously empty cells.
Given an octree cell $C$, let $\centroid_0$ denote its center and $d_0$ denote the UDF value at $\centroid_0$, that is, $d_0 = F(\centroid_0)$. Then we have the following sufficient condition for cell $C$ to be empty: 
 \begin{equation}\label{eq:nec_condition}
    d_0 > \diag(C)/2,
 \end{equation}
where $\diag(C)$ is the diagonal length of cell $C$. Since the UDF distance indicates the distance between a spatial point and the target surface, this criterion determines if the target surface lies outside the sphere centered at $\centroid_0$ with radius $\diag(C)/2$. Given that this sphere contains the entire octree cell, the cell will not contain any portion of the surface if Eqn.~\ref{eq:nec_condition} is satisfied. Otherwise, the cell will first be categorized as an unsolved cell.
Since neural UDF is a reasonable approximation of the ideal UDF, we tailor Eqn.~\ref{eq:nec_condition} to $d_0 > \diag(C)/2 + \epsilon$, where $\epsilon>0$ is a tolerance for the approximation error and set to $\epsilon=2\times 10^{-3}$ throughout all experiments. Empirically, we observed this pruning strategy did not affect the quality of the results.

\textbf{Maximum octree depth. } 
Having a predefined maximal depth of the octree is critical, not only in terms of limiting the amount of computation but also because of the inaccuracy of the neural UDF near the target surface.
Specifically, if the octree keeps subdividing the space that contains the target surface, after a certain depth the cells will become too small and the sample points $\point$ within these cells will lie in the unreliable region of the neural UDF, which will eventually lead to unstable estimation.

\subsection{Point filtering for neural UDFs}\label{sec:adaptation}

Since our method makes use of the gradient directions and the distance values to estimate a surface point in each non-empty grid cell (see Eqn.~\ref{eq:qef}), the quality of the extracted mesh heavily depends on the approximation accuracy of the neural UDF $F(\bf x)$. 
Consider a neighborhood of a sharp feature of the target surface, as shown in the zoom-in views of Fig.~\ref{fig:error_analysis}(e,f). 
The distance errors and gradient direction errors are more significant around the target surface and its cut locus. 

Hence, motivated by our observation and analysis in Sec.~\ref{sec:errors}, to avoid building the QEF using points from these regions, we use the following criteria to filter out unreliable points to enhance the robustness of the QEF solution to the noisy characteristics of the MLP-encoded UDFs.

\noindent 
{\bf Criterion 1: Removing sample points potentially near the surface}.
A candidate point $\point_i$ is considered too close to the target surface if $F(\point_i) < \delta_1$, where $\delta_1$ is a preset filtering threshold; such points are discarded.
This criterion ensures the sample points are from a region where the distance errors are expected to be relatively small.

\noindent
{\bf Criterion 2: Removing sample points whose projections have large UDF values.}
While the previous criterion rejects the sample points that are too close to the target surface and thus avoids the erroneous approximation at those regions, some sample points may still have larger UDF errors even far from the target surface (\eg near the cut locus). 
We observe that these sample points will not be projected to regions near the target surface. Hence, if the UDF of the projected point $\{\proj_i\}$ is large, we consider it to be an incorrect estimate and thus reject the corresponding sample point $\{\point_i\}$ as unreliable.
To this end, we introduce the second filtering criterion as follows, $F(\proj_i) > \delta_2$, where $\delta_2$ is another preset filtering bound; again, the corresponding points $\point_i$ are discarded.

\subsection{Creating Output Mesh Surfaces}

We consider the regular grid of cubic cells at the maximum depth of the octree. Then each edge of the grid is shared by four cubic cells surrounding the edge. To build the initial mesh connectivity, for each edge of the grid, we examine each of its four incident cells. 
Similar to \cite{Ju2002DC}, the mesh connection rule is designed as follows: if all four incident cells are non-empty, a quad-face candidate that connects the four dual points in these cells will be constructed.
To ensure the correct connectivity, we validate and triangulate the quad faces. We further examine if the normals of the face candidates can consistently reflect their geometric property, being a sharp corner, part of a sharp edge, or part of a plane as classified by the SVD shape analysis (Sec.~\ref{sec:qef}). For example, all correct triangulated faces incident to a surface point classified as part of a plane would have all their normals parallel to the singular vector corresponding to the largest singular value from SVD shape analysis, while the normals of triangulated faces incident to a surface point being part of a sharp edge should be orthogonal to its singular vector direction.
This way, we reject all inconsistent face candidates.

We also provide a practical approach to make sure the output mesh is manifold when the desired target surface is manifold:
With the help of the octree cells and the definition of reconstructed dual points, it is feasible to generate an auxiliary \emph{blocky} model by moving all mesh vertices to the centroids of the corresponding cells. By extracting the outer envelope of this auxiliary model, we obtain a manifold structure. 
Finally, we tessellate the surface points using the connectivity from that manifold structure to produce a manifold mesh.
This approach is crucial for applications that require manifold meshes (\eg parameterization, remeshing, or shape analysis).

%% file: sections/results.tex
\section{Experiments}

\subsection{Experimental details and metrics}

\textbf{Experimental setting. }
We rescaled all the shapes to a bounding cube with a side length of 2, centered at the origin. 
To improve efficiency, we share sample points between cells, \ie $m=27$ points in each non-empty leaf cell, including 8 corner points, 12 edge midpoints, 6 face midpoints, and 1 centroid point. We set the filtering parameters $\delta_1=2\times 10^{-3}$ and $\delta_2=2\times 10^{-3}$ to filter the sampled points as described in Sec.~\ref{sec:adaptation}. We notice that more sample points may bring marginal performance improvement but incur computational overhead in our ablation study.

Given that the grid resolution of $128^3$ and $256^3$ are commonly used in related works, we compare our method with these prior techniques using a maximal octree resolution of $128^3$ and $256^3$. 
We also discuss how the resolution will affect our results by varying the maximum depth of the octree to have a resolution from $64^3$ to $512^3$ in our supplementary material.

\begin{figure*}
    \centering
    \includegraphics[width=0.94\textwidth]{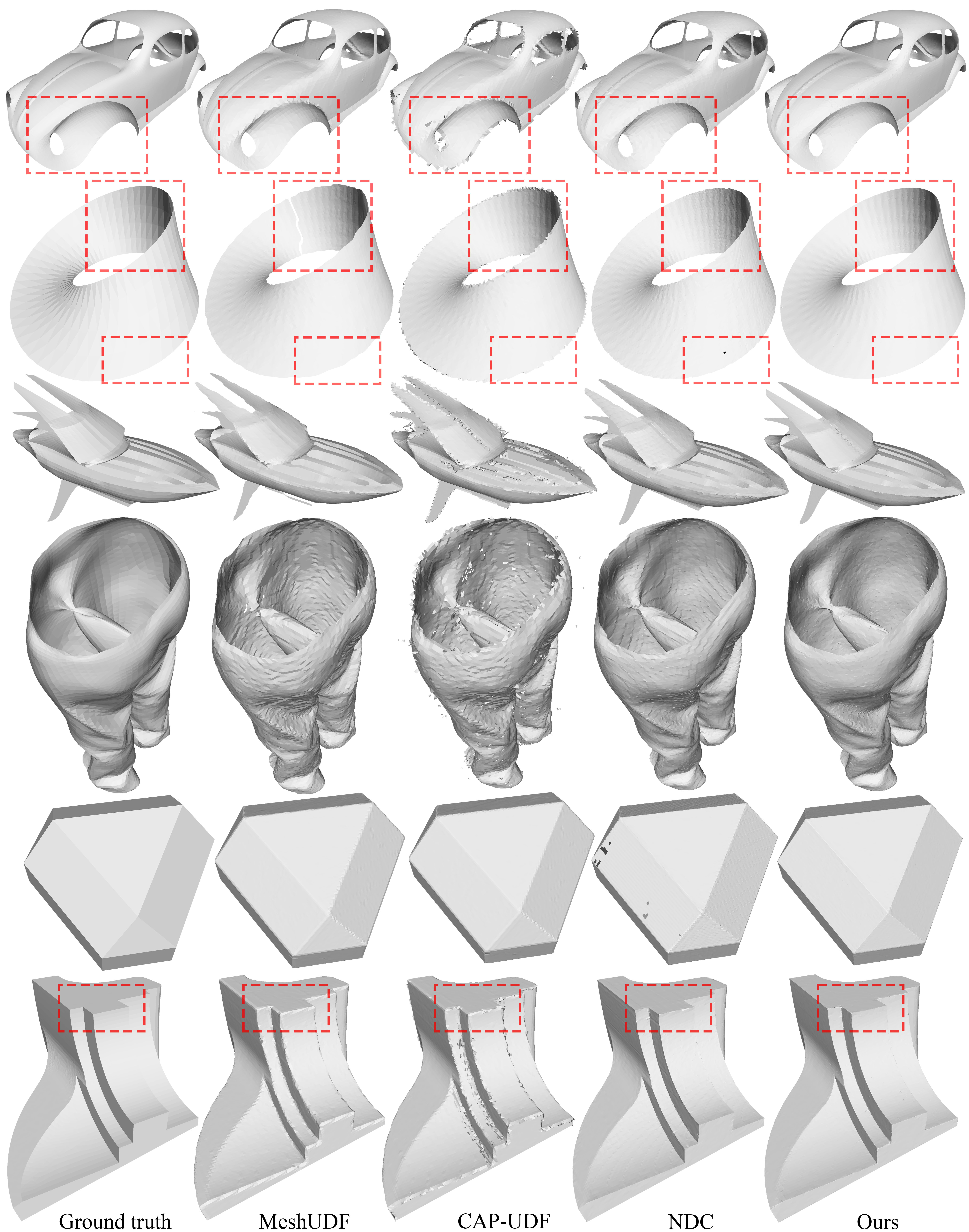}
    \caption{Meshes extracted from neural UDFs. We show 6 examples and compare our results to the mesh extraction results in \textit{MeshUDF}~\cite{Guillard2022MeshUDF}, \textit{CAP-UDF}~\cite{Zhou2022UDF} and \textit{NDC}~\cite{Chen2022NDC}. Our method preserves sharp geometric features better, yields results without undesirable holes, and reconstructs the original smooth boundaries of open surfaces.}
    \label{fig:mlp_gallery}
\end{figure*}
\begin{table*}[htbp]
\centering
\caption{Quantitative comparison between the results obtained by our method and those by the competing methods, MeshUDF, CAP-UDF, and NDC. The average performance on the dataset containing 354 shapes is reported. The Chamfer distance (CD) and the Hausdorff Distance (HD) are scaled by $10^{-4}$ and $10^{-3}$, respectively. T1 and T2 stand for the time spent (seconds) on the mesh extraction and that on the UDF query, respectively.}
\resizebox{0.9\linewidth}{!}{
\begin{tabular}{c | l|| r r r || r r r || r r r || r r}
\hline
& & & MGN & & & Thingi10K & & & ABC & & Running time & \\
& & CD$\downarrow$ & F-score$\uparrow$ & HD $\downarrow$ & CD$\downarrow$ & F-score$\uparrow$ & HD $\downarrow$ & CD$\downarrow$ & F-score$\uparrow$ & HD $\downarrow$ & T1 & T2  \\
\hline \hline
\multirow{4}{*}{$128^3$} & Ours & \textbf{2.38} & \textbf{98.09} & \textbf{11.91} & \textbf{1.97} &  \textbf{97.51}  & \textbf{9.21} & \textbf{3.69} & \textbf{93.41} & \textbf{11.47} & \textbf{0.297} & \textbf{1.184} \\
& MeshUDF~\cite{Guillard2022MeshUDF} & 4.76 & 90.06 & 22.39 & 4.57 & 90.78 & 15.33 &  8.49 & 87.74 & 23.11 & 0.313 & 1.316 \\

& CAP-UDF~\cite{Zhou2022UDF} & 13.53 & 87.10 & 60.04 & 19.55 & 84.85 & 57.49 &  27.75 & 77.43 & 53.46 & 6.466 & 4.764 \\

& NDC~\cite{Chen2022NDC} & 3.32 & 95.61 & 12.04 & 2.74 & 96.44 & 9.69 & 3.95 &92.65 &14.04 &  2.358 & 1.324 \\
\hline \hline

\multirow{4}{*}{$256^3$} & Ours & \textbf{2.03} & \textbf{98.96} & \textbf{7.35} & \textbf{1.60} & \textbf{98.29} & \textbf{6.78} & \textbf{2.22} & \textbf{96.57} & \textbf{8.20} & \textbf{1.492} & \textbf{5.508}  \\
& MeshUDF~\cite{Guillard2022MeshUDF} & 3.45 & 95.10 & 12.8 & 3.33 & 94.13 & 8.91 & 3.25 & 93.85 & 11.49 & 1.600 & 10.27 \\
& CAP-UDF~\cite{Zhou2022UDF} & 8.45 & 92.88 & 49.40 &  12.17 & 90.53 & 47.28 & 21.67  & 83.88 & 43.33 & 49.416 & 37.336 \\
& NDC~\cite{Chen2022NDC} & 2.54 & 98.04 & 7.84 & 2.42 & 97.16 & 7.83  & 2.36 & 96.10 & 11.38 &  9.363 & 10.256 \\

\hline

\end{tabular}}
\label{table:quanti_our_dataset}
\end{table*}

\begin{table}[htbp]
\centering
\caption{Quantitative comparison between our method and three competing methods, MeshUDF, CAP-UDF, and NDC on a pre-trained UDF network with latent codes, on the ShapeNet dataset. In this table, CD and HD are scaled by $10^{-3}$ and $10^{-2}$ respectively. The F-score is calculated with a threshold of $0.006$.
}
\begin{tabular}{l|rrr}
\hline
 & CD$\downarrow$ & F-score$\uparrow$ & HD $\downarrow$  \\
 \hline
Ours & \textbf{3.41} & \textbf{82.30} &  \textbf{3.92} \\
MeshUDF~\cite{Guillard2022MeshUDF} & 4.10 & 79.35 & 4.67 \\
CAP-UDF~\cite{Zhou2022UDF} & 7.70 & 70.30 & 7.21 \\
NDC~\cite{Chen2022NDC} & 7.16 & 78.38 & 12.38 \\
\hline
\end{tabular}
\label{table:shapenet}
\end{table}

\textbf{MLP architecture. }
To overfit single shapes using individual MLPs, we employed the MLP implementation provided by \cite{Sitzmann2020SIREN}. The
activation functions are the Sine activation, except for the
last one which is a \textit{SoftPlus} ($\beta=100$) activation to ensure the output value is non-negative.
All neural UDFs in the experiments were trained with the described MLP implementation. 
All MLP networks were trained for $3k$ iterations with the ADAM optimizer to minimize the difference between the predicted and the GT unsigned distance fields.
For more general tasks, we also test an MLP with latent codes that represent a shape space, following the network and training settings in \cite{Chibane2020UDF}.
We report the timing performance on a Linux desktop with an Intel Core$^\text{TM}$ i7-10870H CPU and an NVIDIA GeForce RTX 3090 graphics card.
We elaborate on the loss function as well as other details regarding training the neural UDFs in the supplementary materials.

\textbf{Metrics.} To evaluate the performance of our method and the other methods, we adopt the following metrics, \ie the double-sided Chamfer distance (CD), the F-score based on CD, and the Hausdorff distance (HD).
The Chamfer distance reflects the overall quality of the extracted surface mesh as compared to the GT shape.
The F-score indicates the percentage of points that are reconstructed correctly under a threshold (set to $0.001$ for shape-overfitting neural UDFs, and $0.01$ for shared UDF network with latent codes).
The Hausdorff distance can reveal if there is anything missing or redundant in the reconstructed geometry (\eg a hole or a floating piece).

\subsection{Comparison with SOTA methods}

\begin{table*}[htbp]
\centering
\caption{Quantitative comparison between two different sampling number settings (27 and 125 per cell respectively). The Chamfer distance (CD) and the Hausdorff Distance (HD) are scaled by $10^{-4}$ and $10^{-3}$, respectively. T1 and T2 stand for the time spent (seconds) on the mesh extraction and that on the UDF query, respectively.}
\resizebox{0.9\linewidth}{!}{
\begin{tabular}{l|| r r r || r r r || r r r || r r}
\hline
 & & MGN & & & Thingi10K & & & ABC & & Running time  & \\
 & CD$\downarrow$ & F-score$\uparrow$ & HD $\downarrow$ & CD$\downarrow$ & F-score$\uparrow$ & HD $\downarrow$ & CD$\downarrow$ & F-score$\uparrow$ & HD $\downarrow$ & T1 & T2  \\
 \hline
 
 27 &   2.38 & 98.09 & 11.91 & 1.97 &  97.51  & 9.21 & \textbf{3.69} & \textbf{93.41} & 11.47 & 0.297 & 1.184 \\
125 & \textbf{2.34} & \textbf{98.11} & \textbf{11.77} & \textbf{1.91} & \textbf{97.52} & \textbf{9.02} & 3.70  & 92.89 & \textbf{11.12} & 0.965 & 7.608 \\

\hline
\end{tabular}}
\label{table:quanti_sampling}
\end{table*}

We compared our results with three existing methods: 1) 
\textbf{MeshUDF}~\cite{Guillard2022MeshUDF}, 2) the \textit{standalone mesh extraction module} presented in \textbf{CAP-UDF}~\cite{Zhou2022UDF}, and 3) the UNDC presented in \textbf{NDC}~\cite{Chen2022NDC}.
The first two methods use gradients of the UDF to estimate sign changes in the field to invoke the Marching Cubes method to extract a surface mesh from the UDF. 
NDC proposes a data-driven approach to extract a surface mesh from the grid-based representation of an implicit field.

\noindent \textbf{Shape-overfitting neural UDFs}

We first compare our results to those obtained by these three methods on a shape collection consisting of shapes from four public sources: 1) 100 from the Thingi10K Dataset \cite{Zhou2016Thingi10k} containing 3D printing models; 2) 134 from the MGN Dataset \cite{Bhatnagar2019MGN} containing garments with open boundaries; 3) 100 from the ABC Dataset \cite{Koch2019ABC} containing CAD models; and 4) 20 commonly used shapes in geometric processing research. 
In addition, we also compare different methods on a M\"{o}bius strip, which is a non-orientable surface. 
We fit each shape with an independent neural UDF and apply mesh extraction to each of these neural UDFs.

Table \ref{table:quanti_our_dataset} reports the performance of different methods on this shape collection. Our method outperforms the other three competing methods in terms of all three quantitative metrics. While all of these methods use uniform grids, our octree structure results in increased computing efficiency. MeshUDF and CAP-UDF would likely be accelerated by adopting an octree structure, however, it would be non-trivial to adopt an octree structure for NDC. For a fair comparison, we also test our method without using the octree acceleration approach for the resolution of $128^3$; by doing so, the T1 time increases to 2.70s, and the T2 time increases to 5.54s.

To qualitatively compare results obtained by different methods, we show the mesh surfaces extracted from neural UDFs for several shapes in Fig.~\ref{fig:mlp_gallery}.
We can see that our results are higher quality than those produced by MeshUDF and CAP-UDF.  
These two methods cannot preserve sharp geometric features due to the use of the Marching Cubes method.
Compared to NDC -- a data-driven method -- our method also produces consistently better results both quantitatively and qualitatively. Specifically, some mesh surfaces extracted by NDC are less smooth than ours. One example (the Pants) is shown in the fourth row in Fig.~\ref{fig:mlp_gallery}. The staircase artifact observed may be attributed to NDC having been trained on the ABC dataset~\cite{Koch2019ABC} containing only mechanical components.

Our method is the only method that recovers open boundaries faithfully, while the other compared methods show staircase artifacts or even redundant pieces near the open boundaries;~see the first two rows of Fig.~\ref{fig:mlp_gallery}.
Also, unexpected holes can be observed on the mesh surface extracted by the competing methods from the neural UDFs. 
Empirically, our method produces quality results without redundant pieces or unexpected holes, generating clear-cut boundaries of the open surfaces as shown in the top four rows of Fig.~\ref{fig:mlp_gallery}, and reproducing the sharp features as shown in the bottom two rows of the same figure.

More results and comparisons (also on the GT UDF) are presented in our Supplementary Materials, which further validate the superior performance of our method.

\textbf{Shared, \textit{pre-trained} UDF network with latent codes.}
We consider a single neural network with latent codes trained to represent the entire shape space, where each shape is associated with a unique latent code.
Specifically, we test our method and the competing methods on the pre-trained UDF network provided by \cite{Chibane2020UDF} representing 300 ShapeNet car models. Although this UDF network is less accurate than the other overfitting-based network settings considered earlier, our method still outperforms the other methods by a significant margin as reported in Tab.~\ref{table:shapenet}, showing the versatility and robustness of our method.

\subsection{Ablation study}
We conducted an ablation study on our method using sampling numbers (the number $m$ in Eqn.~\ref{eq:qef}) of 27 ($3^3$) and 125 ($5^3$) to justify our design choice to use 27 as our sampling number.
In Tab.~\ref{table:quanti_sampling}, we demonstrate that using 125 sample points per cell results in a marginal improvement but incurs a significantly higher computing cost.

\subsection{Limitations}
While DualMesh-UDF demonstrates the ability to accurately extract surfaces from neural UDFs, some limitations remain. 
First, the extracted surface cannot be adaptively subdivided with respect to sharp features or fine geometry details. 
To faithfully reconstruct the fine geometry details, we need to pre-define a sufficient depth of the octree and solve the QEF problem in each of its non-empty leaf nodes.
Second, to share the UDF and gradient values at sampled points between adjacent octree cells, we adopt a regular grid sampling strategy. A more flexible and adaptive sampling strategy may bring further improvement to our method.

%% file: sections/conclusion.tex
\section{Conclusion}

We have presented a method for extracting high-quality surface meshes from unsigned distance fields. In order to attain robust performance on MLP-encoded neural UDFs, we discuss the characteristics of the approximation errors of the neural UDF and develop an adaptive octree-based method to effectively localize the target surface embedded in the given UDF. Extensive experiments show that our method outperforms the SOTA methods and produces high-quality results with sharp geometry features, with clear open boundaries, and free of undesirable holes.

\section{Acknowledgement}
This research is partly supported by the Innovation and Technology Commission of the HKSAR Government through the InnoHK initiative. The authors would like to thank Yanhong Lin and Ruixing Jia for their generous help on the project.

%% file: supp_sections/MLP_characteristics.tex
\section{Error characteristics of MLP-encoded UDF in 3D}
To supplement the discussions of Sec.~3 in the main text, in this section,  we provide more information about the error characteristics of a neural UDF in the 3D case, which is most relevant to the setting of our algorithm.

\begin{figure*}[htbp]
    \centering
    \includegraphics[width=0.95\linewidth]{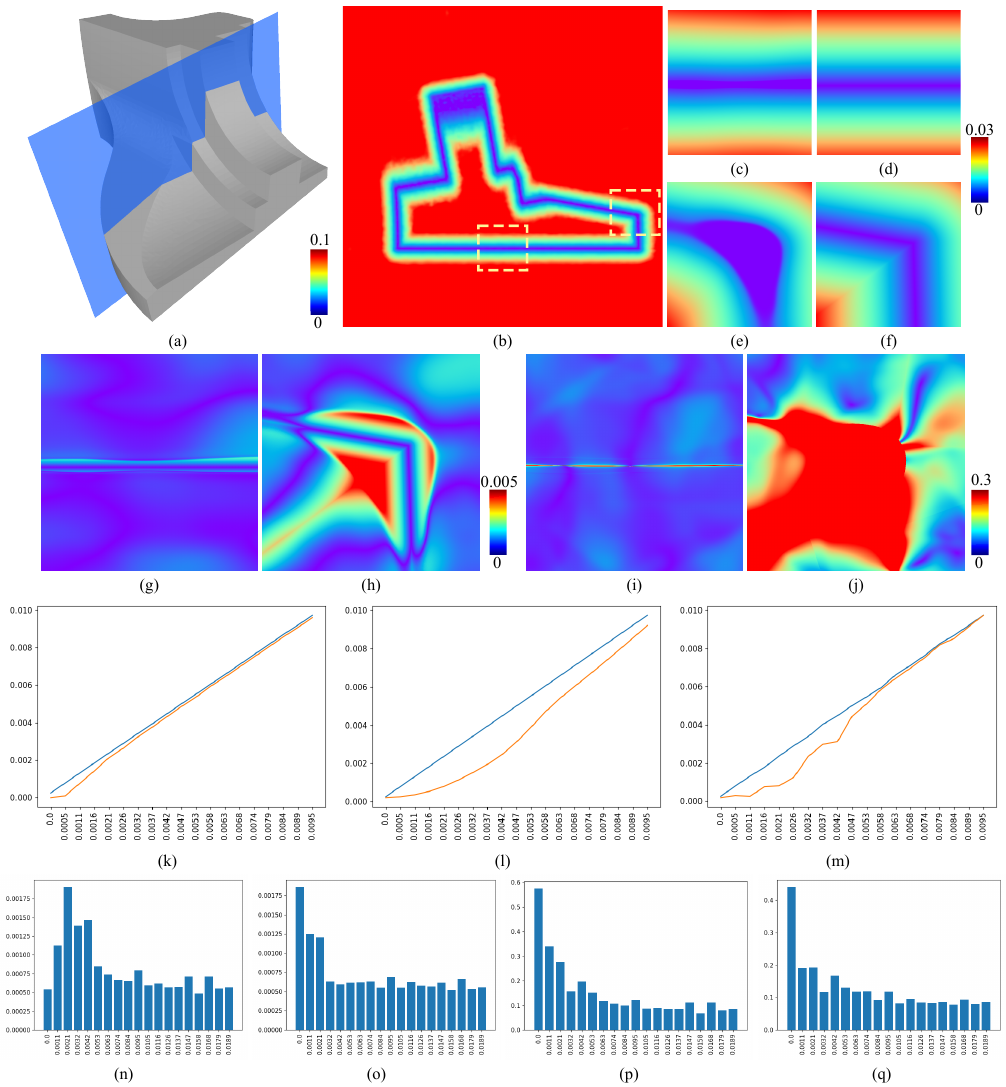}
    \caption{The errors of a neural UDF on \textit{Fandisk}. {\bf (a)} A plane cutting across the Fandisk model; {\bf (b)} The color map of an MLP-encoded UDF of the Fandisk on the cross-section in (a); {\bf (c)-(f)} The closeup views of the neural UDF ((c), (e)) and the corresponding GT UDF ((d), (f)) in the two framed local regions shown in (b), respectively; {\bf (g)-(j)} The color maps of the errors of the neural UDF ((g), (h)) and the errors of the neural UDF's {\em normalized} gradient vector fields in the same two local regions ((i), (j)), respectively; {\bf (k)-(m)} The plots of the averages of the GT UDF values (blue curve) and the neural UDF values (orange curve) in the same two local regions (around plane (k), around edge (l)), and around the overall surface (m), respectively, with the GT UDF value being the horizontal axess;  {\bf (n)-(o)} The histogram of the errors of the neural UDF (n) and the histogram the errors of normalized gradient vector fields (o), respectively, with {\em the GT UDF values} being the horizontal axes.  {\bf (p)-(q)} The histogram of the errors of the neural UDF and the histogram of the errors of normalized gradient vector fields, respectively, with {\em the neural UDF values} being the horizontal axes.}
\label{fig:fandisk_error}
\end{figure*}

Fig.~\ref{fig:fandisk_error}(b) shows the GT UDF of the Fandisk model at a cross-section (Fig.~\ref{fig:fandisk_error}(a)). 
Fig.~\ref{fig:fandisk_error}(c) and (d) show the close-up views of the neural UDF values and the GT UDF values of the Fandisk model around a flat region, and Fig.~\ref{fig:fandisk_error}(e) and (f) show another comparison between the neural and GT UDF values around a sharp edge.
Fig.~\ref{fig:fandisk_error}(g)-(j) present the color-coded error maps of the neural UDF values ((g) and (h)) and of the normalized gradient directions of the neural UDF ((i) and (j)), for the same two local regions. 
The GT UDF values (blue curves) and the neural UDF values (orange curves) are plotted regarding the same flat region in Fig.~\ref{fig:fandisk_error}(k), regarding the same sharp edge in Fig.~\ref{fig:fandisk_error}(l), and regarding the entire surface in Fig.~\ref{fig:fandisk_error}(m).
Histograms are plotted to show the overall distributions of the neural UDF errors in Fig.~\ref{fig:fandisk_error}(n) and its gradient vector errors in Fig.~\ref{fig:fandisk_error}(o), respectively, against the GT UDF values (the horizontal axes). Finally, in Fig.~\ref{fig:fandisk_error}(p) and (q), we visualize the distributions against neural UDF values being the horizontal axes.

The observations about the error characteristics of the neural UDF made in this 3D case align well with the 2D case provided in the main text, which are:
\begin{enumerate}
    \item[(1)]  The errors of the neural UDF are concentrated around the target surface, \ie, the zero-level set of the ideal UDF;
    \item[(2)]  The errors around the sharp edges are more pronounced than those around a flat region; and
    \item[(3)]  The errors of the neural UDF are too large to be reliable within the distance of 0.002 from the surface around a flat region and within the distance of 0.005 from the surface around a sharp edge. \footnote{All distance thresholds are given based on shapes normalized to a bounding region of $[-1,1]^3$.}
\end{enumerate}

In our algorithm, we choose the sampling threshold to be 0.002, based on the overall error characteristics as reflected in the histograms in Fig.~\ref{fig:fandisk_error}(p) and (q). That is no sampling points are selected with their neural UDF values less than 0.002.
While the qualitative observations (1) and (2) hold generally for all kinds of neural UDFs, the numerical characterization in observation (3) is specific to the particular MLP architecture and the training strategy used in our experimental setup. Different MLP architectures or training strategies might lead to more or less accurate neural UDFs, as we will show later in this Supplementary Materials.

%% file: supp_sections/training_settings.tex
\section{MLP-encoded UDF training details}
We will introduce the loss function, sampling strategy, and more detailed training configuration in this section.

The overfitting network that we used in the main text is a 9-layer MLP with each layer having 512 neurons.
The activation functions are the Sine activation, except for the last one which is a SoftPlus ($\beta=100$) activation to ensure output values are non-negative.

\textbf{Loss function.} The loss function to train an overfitting network is presented as follows:
\begin{equation}
    \mathcal{L} = \int_{\Omega } |f_{\theta}(\mathbf{x}) - \mathrm{UDF}(\mathbf{x})| d\mathbf{x}
\end{equation}
where $f_{\theta}(\mathbf{x})$ and $\mathrm{UDF}(\mathbf{x})$ denote the neural and GT unsigned distance values at point $\mathbf{x}$, respectively.

\textbf{Sampling strategy. }
For each shape, we normalize the shape so that it is centered at the origin and bounded in a domain $\Omega = [-1,1]^3$. The training samples of each shape consist of 600k points uniformly sampled on the shape surface, 1200k points within the distance of 0.05, 800k points within the distance of 0.3 following Gaussian distribution, and 400k points uniformly sampled in the bounding domain $\Omega$.

\textbf{Training configuration. }
The mini-batch size is set to 30k samples. The initial learning rate is set to $1\times 10^{-4}$ and is decayed by $0.3$ after 1500 and 2300 iterations.

%% file: supp_sections/experiments.tex
\section{Supplementary experiments}

We present additional experiments and discussions on 1) visualizing the mesh extraction results of different methods on GT UDFs; 2) performance on neural UDFs trained with another frequently used setting, \ie using the positional encoding and replacing ReLU activation layers with Softplus; 3) our method under different grid resolutions; and 4) showcases of our reconstructed mesh quality.

\subsection{Evaluation on ground-truth UDFs}

\begin{figure*}
    \centering
    \includegraphics[width=0.95\linewidth]{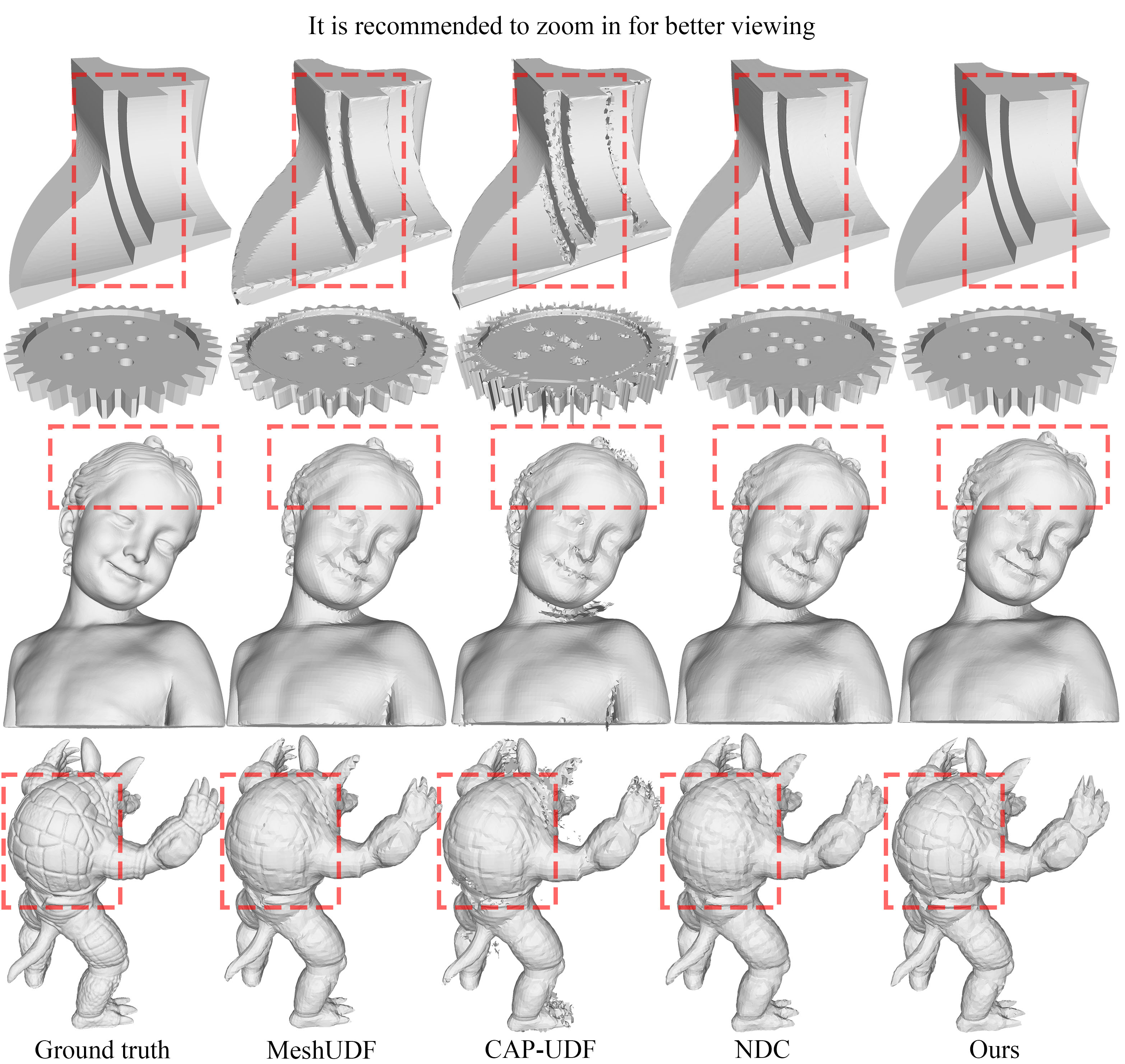}
    \caption{Meshes extracted from ground truth UDFs. We show four examples and compare our results to MeshUDF \cite{Guillard2022MeshUDF}, CAP-UDF \cite{Zhou2022UDF}, and NDC \cite{Chen2022NDC}. Our method preserves sharp geometric features better (top two rows), and yields smoother results on organic models (bottom two rows).
    }
    \label{fig:gt_gallery}
\end{figure*}

We evaluate our method, MeshUDF \cite{Guillard2022MeshUDF}, CAP-UDF \cite{Zhou2022UDF}, and NDC \cite{Chen2022NDC} on a set of ground truth UDFs that are directly generated by 3D meshes. In Fig.~\ref{fig:gt_gallery}, we present four models for visual comparison. Our method preserves both sharpness and smoothness more faithfully when compared to the other three methods.

\subsection{Generalization to MLPs trained with different settings}

\begin{figure*}
    \centering
    \includegraphics[width=0.89\linewidth]{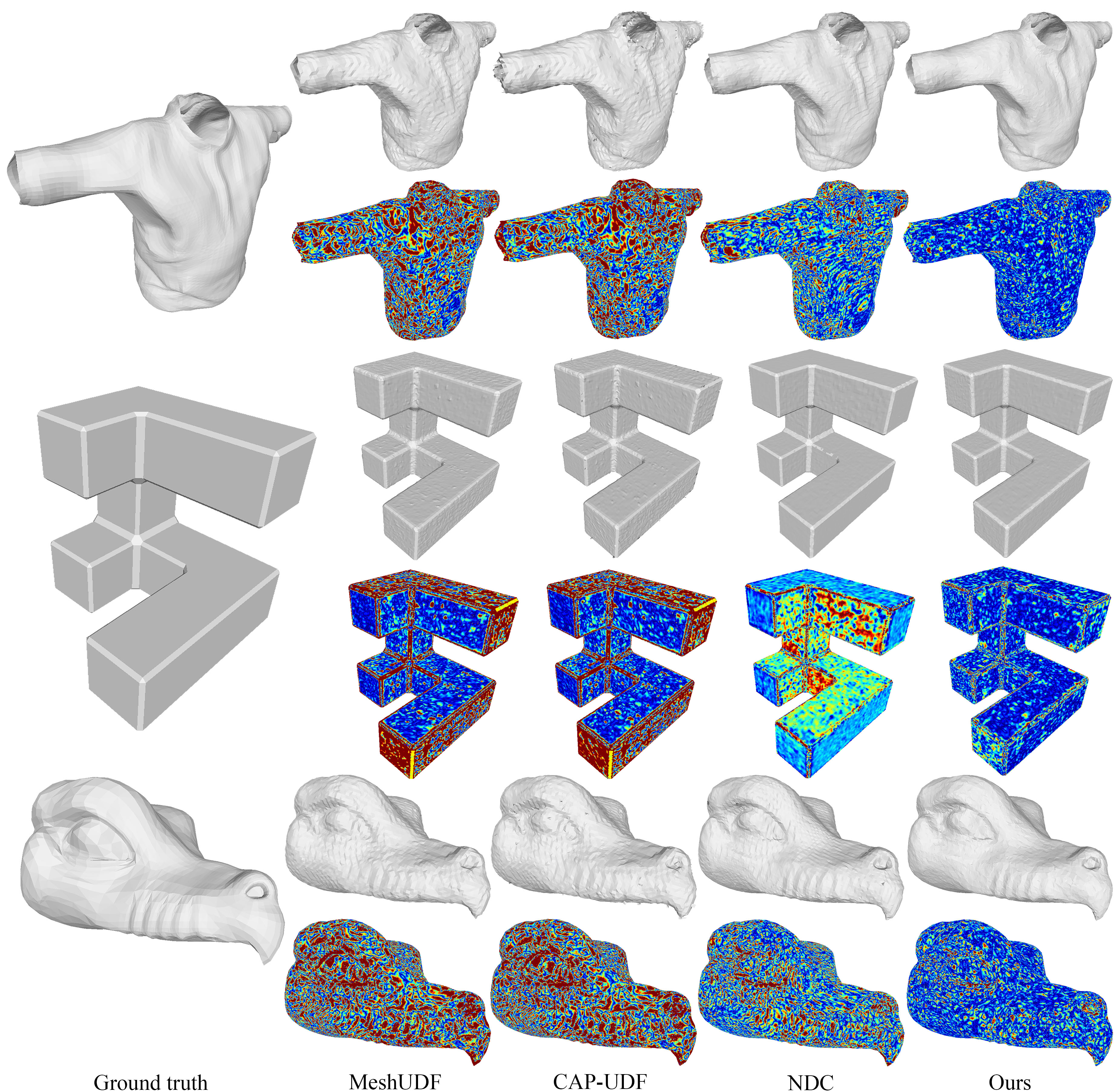}
    \caption{Comparisons of extracted meshes and the corresponding color-coded error maps obtained by our method and the competing methods. The warmer color indicates a larger error. Our method consistently outperforms the competing methods in all examples, achieving lower errors, preserving sharp features, and reproducing smooth surfaces and geometric details. MeshUDF and CAP-UDF cannot preserve the sharp features of the shapes (the 3rd row). The three competing methods find it difficult to cope with smooth transitions in the surface, yielding staircase artifacts.}
    \label{fig:gallery_with_error_maps}
\end{figure*}

\begin{table*}[htbp]
\centering
\caption{Quantitative comparison between the results obtained by our method and those by the competing methods, \ie MeshUDF, CAP-UDF, and NDC. Different from Table \ref{table:quanti_our_dataset}, this experiment follows MeshUDF \cite{Guillard2022MeshUDF} to use the positional encoding but replace all ReLU activations with SoftPlus activations. The average performance on the dataset containing 354 shapes is reported. The Chamfer distance (CD) and the Hausdorff Distance (HD) are scaled by $10^{-4}$ and $10^{-3}$, respectively.}
\resizebox{0.9\linewidth}{!}{
\begin{tabular}{l|| r r r || r r r || r r r }
\hline
 & & MGN & & & Thingi10K & & & ABC & \\
 & CD$\downarrow$ & F-score$\uparrow$ & HD $\downarrow$ & CD$\downarrow$ & F-score$\uparrow$ & HD $\downarrow$ & CD$\downarrow$ & F-score$\uparrow$ & HD $\downarrow$  \\
 \hline
Ours & \textbf{4.25} & \textbf{87.95} & \textbf{10.75} & \textbf{4.58} & \textbf{88.24} & \textbf{13.99} & \textbf{5.19} & \textbf{88.23} & \textbf{10.63}  \\
MeshUDF~\cite{Guillard2022MeshUDF} & 11.31 & 54.10 & 17.99 & 10.46 & 58.66 & 17.65 & 13.47 & 68.63 & 17.12 \\
CAP-UDF~\cite{Zhou2022UDF} & 18.22 & 52.09 &37.78 & 19.98 & 53.47 & 34.88 & 35.49 &  62.98 & 37.20   \\
NDC~\cite{Chen2022NDC} &6.26 & 73.24 & 11.09 & 6.92 & 72.54 & 16.77 & 7.45 & 80.29 & 15.04   \\
\hline
\end{tabular}}
\label{table:quanti_our_dataset_pe}
\end{table*}

\begin{figure*}
    \centering
    \includegraphics[width=\linewidth]{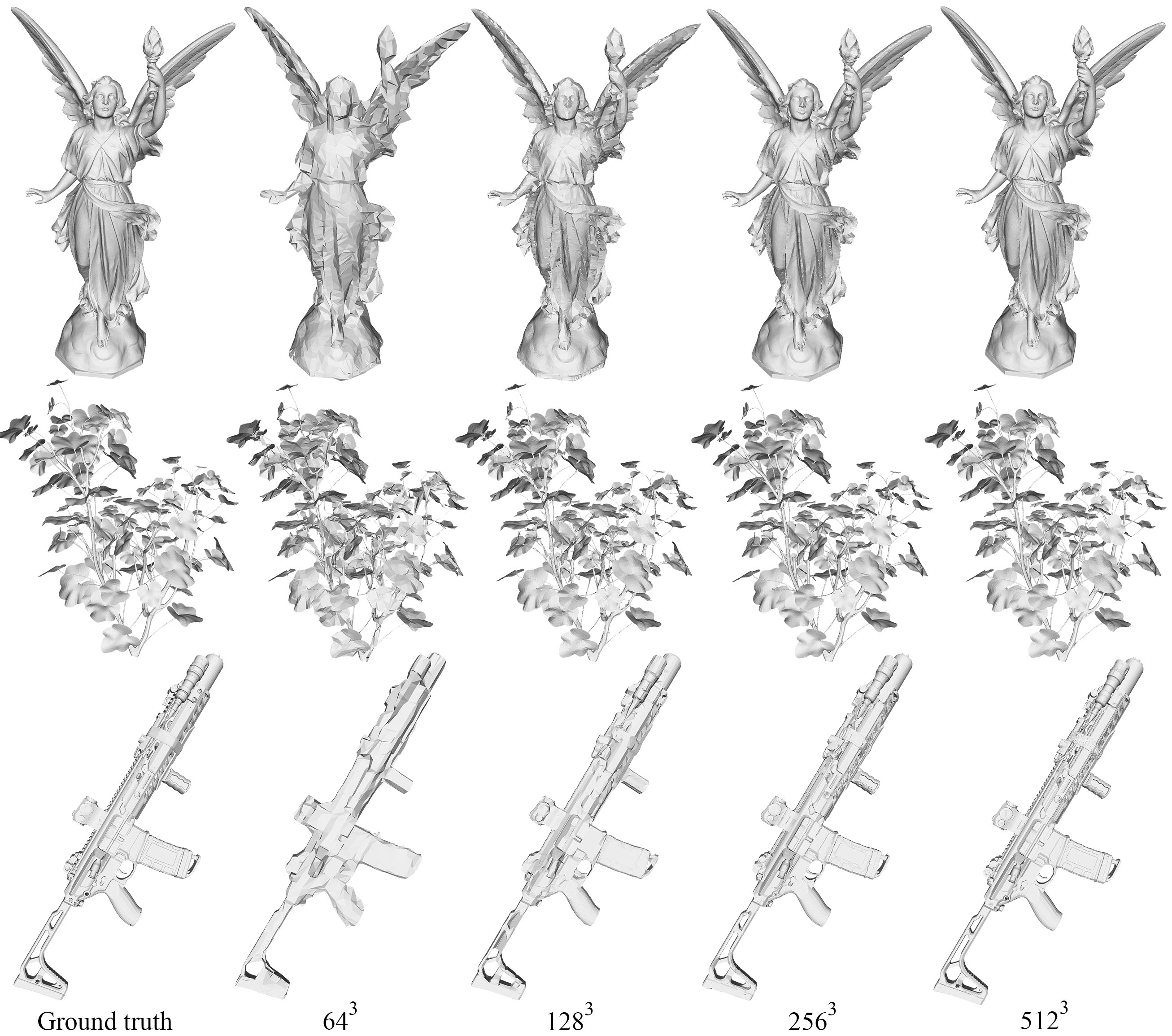}
    \caption{Extracted meshes from GT UDFs obtained by our method with different grid resolutions.}
    \label{fig:multi_resolution_on_gt}
\end{figure*}

\begin{figure*}[htbp]
    \centering
    \includegraphics[width=0.99\linewidth]{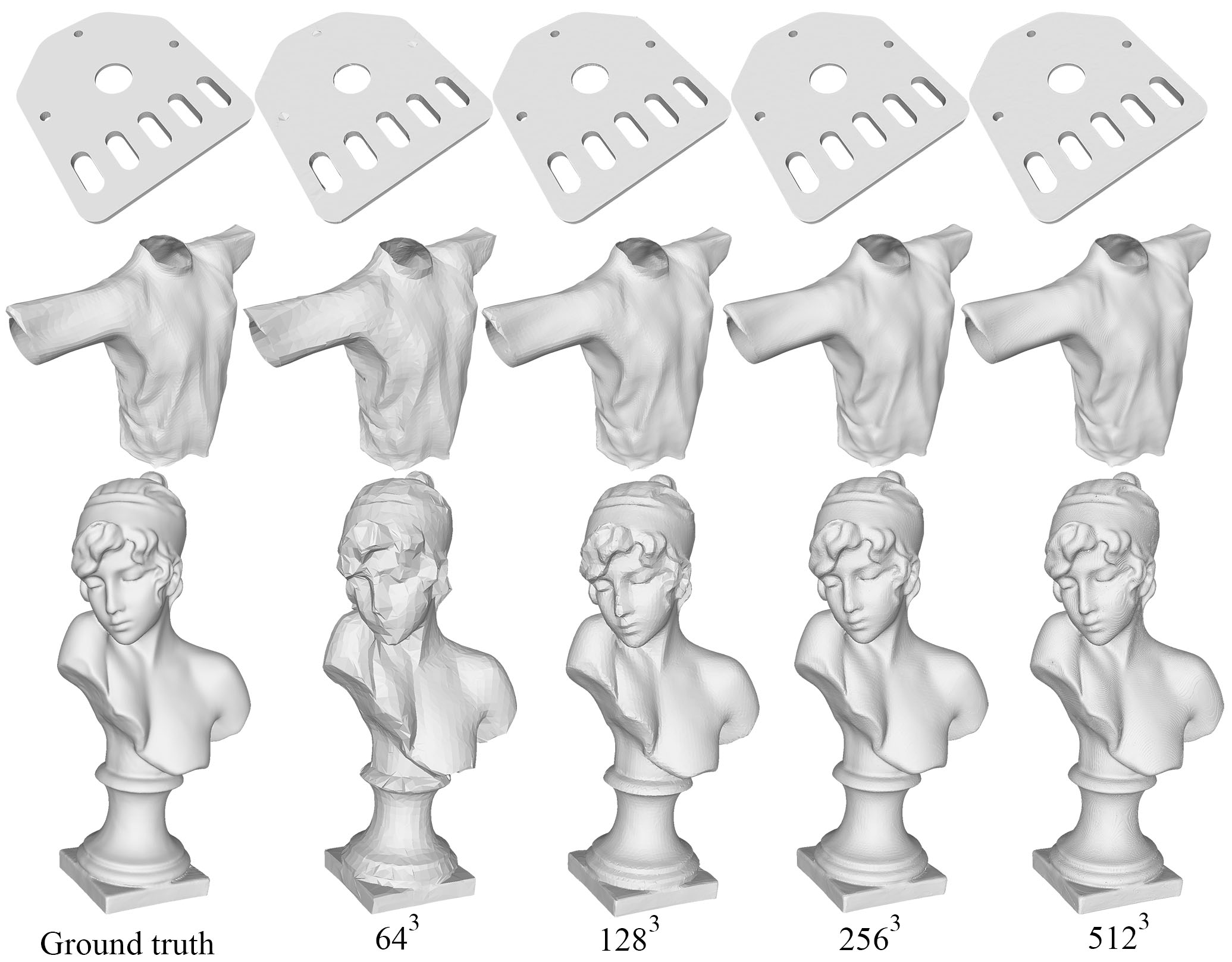}
    \caption{Extracted meshes from neural UDFs obtained by our method with different grid resolutions.}
    \label{fig:multi_resolution_on_mlp}
\end{figure*}

\begin{figure*}[htbp]
    \centering
    \begin{subfigure}{0.35\linewidth}
    \includegraphics[width=\linewidth]{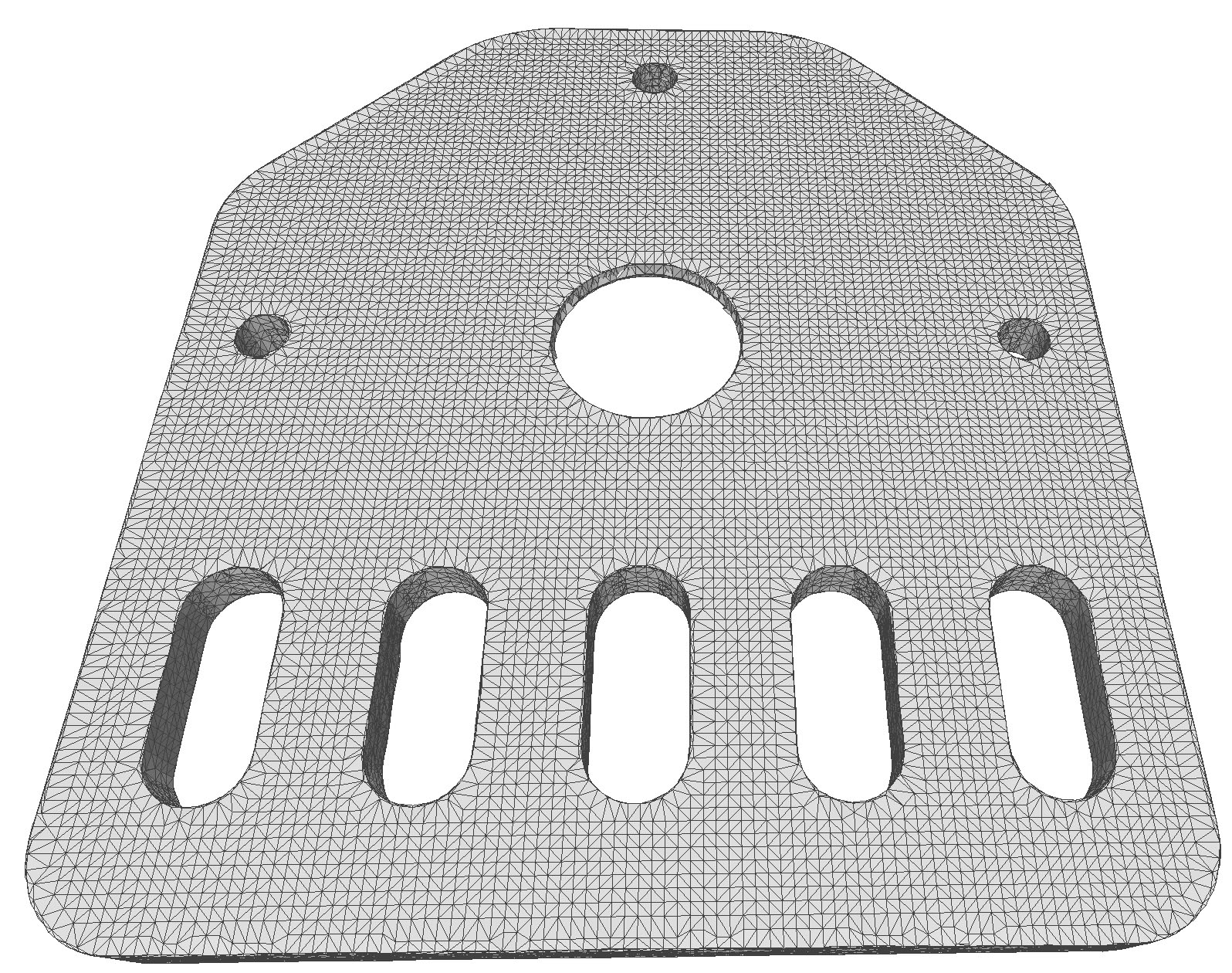}
    \label{fig:meshudf_relu}
    \end{subfigure}
    \begin{subfigure}{0.35\linewidth}
    \includegraphics[width=\linewidth]{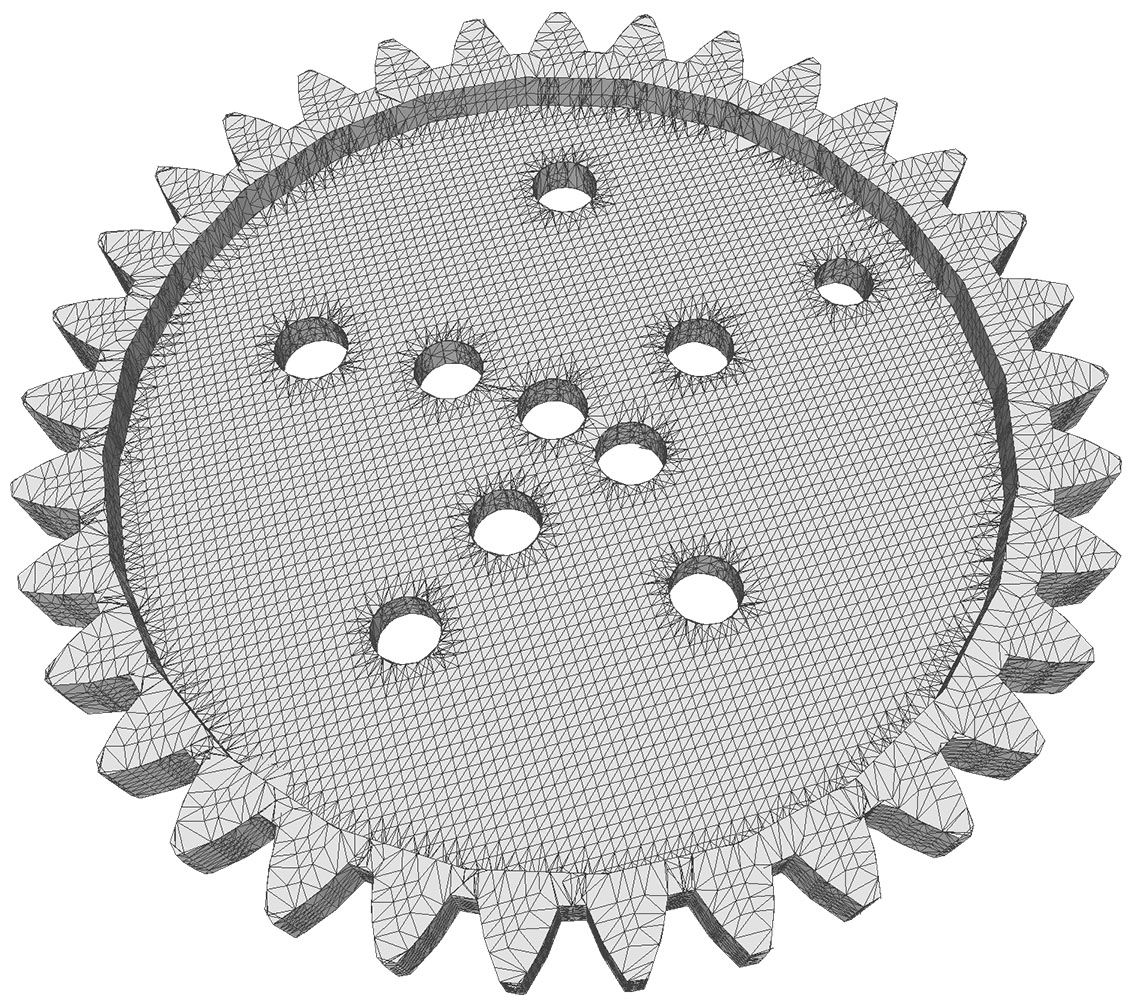}
    \label{fig:ndc_relu}
    \end{subfigure}
    \caption{The visualization of the mesh quality of our reconstruction results.}
    \label{fig:mesh_connectivity}
\end{figure*}

\begin{table*}[htbp]
\centering
\caption{Quantitative results of $64^3$, $128^3$, $256^3$, and $512^3$ resolutions. The Chamfer distance (CD) and the Hausdorff Distance (HD) are scaled by $10^{-4}$ and $10^{-3}$, respectively. T1 and T2 stand for the time spent (seconds) on the mesh extraction and that on the UDF query, respectively.}
\resizebox{0.9\linewidth}{!}{
\begin{tabular}{l|| r r r || r r r || r r r || r r }
\hline
 & & MGN & & & Thingi10K & & & ABC & & & Running Time \\
 & CD$\downarrow$ & F-score$\uparrow$ & HD $\downarrow$ & CD$\downarrow$ & F-score$\uparrow$ & HD $\downarrow$ & CD$\downarrow$ & F-score$\uparrow$ & HD $\downarrow$ & T1$\downarrow$ & T2$\downarrow$ \\
 \hline
$64^3$ & 5.55 & 86.23 & 23.56 & 5.70 & 90.57 & 19.18 & 13.54 & 81.99 & 23.20 & 0.098 & 0.324  \\
$128^3$ & 2.38 & 98.09 & 11.91 & 1.97 & 97.51  & 9.21 & 3.69 & 93.41 & 11.47 & 0.297 & 1.184 \\
$256^3$ & \textbf{2.03} & \textbf{98.96} & 7.35 & \textbf{1.60} & 98.29 & \textbf{6.78} & 2.22 & \textbf{96.57} & \textbf{8.20} & 1.492 & 5.508 \\
$512^3$ & 2.28 & 98.48 & \textbf{7.01} & 1.74 & \textbf{98.88} & 6.91 & \textbf{2.00} & 95.68 & 9.22 & 9.362 & 28.56 \\
\hline
\end{tabular}}
\label{table:quanti_multi_resolution}
\end{table*}

We have reported all quantitative comparisons in the main text using the Sine activation~\cite{Sitzmann2020SIREN} for the hidden layers.
In this supplementary material, we also tested our method on neural UDFs learned with other network settings
and compared our results with those of the other three competing methods. 
Specifically, we follow the setting in MeshUDF~\cite{Guillard2022MeshUDF} where the positional encoding is used before passing the coordinate queries into the network, and all activations are replaced with Softplus ($\beta=100$). The hyperparameters of the proposed are unchanged during this experiment.

We compare our method with MeshUDF \cite{Guillard2022MeshUDF}, CAP-UDF \cite{Zhou2022UDF}, and NDC \cite{Chen2022NDC} on our shape-overfitting neural UDF dataset including 354 shapes in total as we introduced in our main text. These comparisons show that our results are superior to all three methods in terms of approximation errors, visual smoothness, and preservation of sharp edges and smooth surface boundary curves.
This validates the applicability of the proposed method to neural UDFs trained with other network settings.

Specifically, in Fig.~\ref{fig:gallery_with_error_maps}, we visualize the distribution of approximation errors over the reconstructed meshes, measured by the distances from the GT mesh to the reconstructed mesh in each case.
All the color-coded error maps are ranged in $[0, 0.0015]$ (from cool to warm) with errors larger than $0.0015$ clamped at $0.0015$.
The results validate that our method produces smaller reconstruction errors overall, especially around the geometric features, than the other methods.

\subsection{Experiment on different grid resolutions}

We validate our method's capability at a wide range of resolutions.
Since other competing methods are based on regular grids, evaluating them on finer resolutions will exceed our memory limit.
In this experiment, we test our method under different octree max depths (\ie different grid resolutions) on several complex shapes for GT UDF and our neural UDF datasets that we introduced in our main text, respectively. 

We show the visual comparison in Fig.~\ref{fig:multi_resolution_on_gt} for GT UDFs and Fig.~\ref{fig:multi_resolution_on_mlp} for neural UDFs, respectively. It is recommended to zoom in for better viewing.
In Table.~\ref{table:quanti_multi_resolution}, we report the quantitative results of $64^3$, $128^3$, $256^3$, and $512^3$ resolutions on the neural UDF datasets. 
We note that the edge length of a cell under the grid resolution of $512^3$ is around $0.004$, which is twice the threshold $\delta_1$ of $2\times 10^{-3}$ set for sample filtering in Sec.~4, hence there may be insufficient samples (i.e., less than 3 valid sample points) within this cell after filtering. To address this issue, whenever a cell at this resolution level has less than 3 valid sample points, we lower $\delta_1$ to $0.001$ to collect enough samples to solve the QEF problem and estimate the surface point in this cell. As we mentioned in our main text, when cells become too small, the sample points within these cells will lie in the unreliable region of the neural UDF, which will eventually lead to unstable estimation.

\subsection{Mesh quality demonstration}

Our surface point positioning strategy and mesh surface generation method result in satisfactory mesh quality. The mesh connectivity of two of our results is visualized in Fig.~\ref{fig:mesh_connectivity}.